\newcommand {\be}{\begin{equation}}
\newcommand {\ee}{\end{equation}}
\newcommand {\bea}{\begin{eqnarray}}
\newcommand {\eea}{\end{eqnarray}}
\newcommand  {\om}{K}
\newcommand {\bom} {{ \bf K}}
\newcommand {\bo} {{\bf W}}
\newcommand {\ide} {{\bf 1}}
\newcommand {\rat} {{ \bar \om_{max}/\bar \om_{min}}}

\documentstyle[preprint,aps,prb]{revtex}
\begin{document}
\draft
\title{ Acceleration Schemes for Ab-Initio Molecular Dynamics and Electronic
Structure Calculations
}
\author {F. Tassone$^{a,c}$ F. Mauri$^{a,b}$   and R. Car$^{a,b}$}
\address{
(a) Institut Romand de Recherche Numerique en Physique
des Materiaux (IRRMA) \\
PHB-Ecublens, CH-1015 Lausanne ,Switzerland\\
(b)  Dept. of Condensed Matter Physics, University of Geneva,\\
CH-1211 Geneva, Switzerland\\
(c) Scuola Normale Superiore, P.za dei Cavalieri 7, I-56126
Pisa,Italy\\}
\date{\today} \maketitle
\begin{abstract}
We study the convergence and the stability of fictitious dynamical
methods for electrons.
First, we show that a particular damped second-order
dynamics has a much faster rate of convergence to the ground-state
than first-order steepest descent algorithms while retaining their
numerical cost per time step.
Our damped dynamics has efficiency comparable to that of
conjugate gradient methods in typical electronic minimization problems.
Then, we analyse the factors that limit the size of
the integration time step in approaches based on plane-wave expansions.
The maximum allowed time step is dictated by the highest
frequency components of the fictitious electronic dynamics.
These can result either from the large
wavevector components of the kinetic energy or from the small wavevector
components of the Coulomb potential giving rise to the so called {\it charge
sloshing} problem. We show how to eliminate large wavevector instabilities
by adopting a preconditioning scheme that is implemented here for the
first-time in the context of Car-Parrinello ab-initio molecular dynamics
simulations of the ionic motion. We also show how to solve
the charge-sloshing problem when this is present.
We substantiate our theoretical analysis with numerical tests
on a number of different silicon and carbon systems having both insulating and
metallic character.
\end{abstract}

\section{Introduction}

The introduction of a fictitious dynamics for the electrons\cite{Car,lh} with
driving forces obtained from the total energy within
density functional theory (DFT)\cite{DFT} has provided a convenient approach to
minimize the total energy of condensed matter systems
and to perform ab-initio molecular dynamics simulations of the ionic motion.
These techniques
have been applied successfully to a variety of insulating, semiconducting,
and metallic systems involving a large number of atoms in the context of
structural optimization problems at zero temperature and of dynamical
simulations of the atomic motion at finite temperature \cite{review}.

It is a subject of current interest to study the factors that limit
the efficiency of fictitious dynamical methods for electrons in order
to improve their numerical efficiency.
This depends on the
choice made for the dynamics and on the size of the time step that
can be used to integrate numerically the equations of motion.

When discussing how to choose a specific dynamics,
it is convenient to consider total energy minimization separately from
molecular dynamics. It has been shown by Car and Parrinello that to
simulate the classical adiabatic motion of the atoms it is useful to adopt a
second-order Newtonian dynamics also for the electronic degrees of freedom,
since this exploits optimally the concept of continuous simultaneous
evolution of electronic and atomic degrees of freedom\cite{Car,lh}. Newtonian
dynamics conserves energy. Different approaches should be used
to minimize the electronic energy, as it is required to start a molecular
dynamics simulation or to solve an optimization problem at zero temperature.
The simplest approach to minimization is provided by steepest-descent dynamics,
which can be viewed as a dynamics of the first-order in the
time-derivative\cite{lh}.
Steepest-descent dynamics, which requires only knowledge of the gradients of
the
energy functional, is not very efficient particularly in metallic
situations. Better schemes require some knowledge also of the second
derivatives of the energy functional either explicitly or implicitly.
Conjugate gradient methods have been
developed in this context\cite{numerical,Stich,Teter,Arias} and have been shown
to be
superior
to steepest-descent methods, particularly when the full energy functional
was used in the line minimizations and full account was taken of
the orthonormality constraints on the wavefunctions\cite{Arias}.

In this paper we show that a minor modification of a steepest-descent
algorithm, namely replacing first-order dynamics with a specific
damped second-order dynamics, improves substantially the rate of convergence
of the wavefunctions to the ground-state. The
resulting scheme, which we call damped
molecular dynamics, has efficiency comparable to that of the best
conjugate gradient algorithms when used in typical electronic minimization
problems, with the additional advantage of having basically the same numerical
complexity of simple steepest-descent algorithms.

We then investigate what determines the maximum allowed time step for
numerical integration when using
steepest-descent (SD), damped (D) or Newtonian molecular dynamics (MD).
In all cases the time step is limited by the need to integrate
the high frequency components of
the fictitious dynamics. These
arise either from the large wavevector components of the electronic kinetic
energy or from the small wavevector components of the Hartree energy due
to the divergence of the Coulomb potential at small wavevector.
In the latter case the related numerical instability is usually
referred to as the `charge sloshing' problem and it is expected to
become serious when the size of the system becomes very large.

The large wavevector instability can be eliminated by preconditioning
the equations of motion since at large wavevectors
the wavevefunctions are dominated by the kinetic energy and
are to a large extent free-particle like. Indeed, it was already
suggested earlier by several authors that this property could be used to
speedup
iterative schemes for electronic minimization. In particular,
Ref.~(\onlinecite{Payne}) proposed
an analytical integration scheme for the large wavevector components of the
wavefunctions within second-order dynamics.
This scheme was subsequently extended
in Ref.~(\onlinecite{Soler}) to first-order steepest-descent equations.
Since this
approach can be less stable
than standard steepest-descent algorithms\cite{Stich} we will not
discuss it any further.
So far the most successful preconditioning scheme in the context of electronic
minimization is the one proposed in
Ref.~(\onlinecite{Teter})
in connection with a conjugate-gradient method to minimize the
total energy.

In this paper, we propose
a preconditioning scheme which is appropriate to all the
dynamical methods referred to above, namely SD, D, and MD dynamics.  It
consists in properly scaling the
fictitious masses associated to the large wavevector components
of the electronic wavefunctions, in order to compress the high frequency
spectrum of the electronic dynamics and to use a larger integration time
step. Our preconditioning method
is similar in spirit to the one described in Ref.~(\onlinecite{Teter})
in the context of conjugate gradient minimization but it is formulated
as a modification of the differential equations leading to SD, D, and
MD dynamics. In particular, we apply it here for the first time
to the Car-Parrinello MD equations, which provide an
efficient approach for ab-initio molecular dynamics
simulations of the ionic motion. In this context our preconditioning scheme
allows to use a timestep which is two to three times larger than in previous
applications of this method, resulting in a considerable saving of
computational time.

We now turn our attention to the `charge sloshing' problem. This
has been discussed previously in the context of self-consistent
diagonalization of the
Kohn-Sham Hamiltonian\cite{Martin}.
The onset of this kind of instability is expected to
occur at
significantly larger sizes in the context of fictitious dynamical methods since
in these approaches
the wavefunctions change little over a single timestep. Indeed recent MD
simulations for metallic liquid silicon
have shown {\it no sign}
of a `sloshing' instability up to cubic cells containing 216 silicon
atoms\cite{Sugino}.
However one expects that for sufficiently large cells the `sloshing'
instability
should appear, although a quantitative theoretical analysis of it
in the context of fictitious dynamical methods for electrons has been so far
missing. `Sloshing' instabilities have been found numerically within
some iterative schemes for electronic minimization in the case of systems
having a very long linear dimension\cite{Teter}.
In this paper we present a theoretical analysis of the `charge
sloshing' problem in the context of SD, D, and MD equations of motion.
We find that the `sloshing' instability is absent for insulators,
but it is present for metals. This is in accord with previous results of
Ref.~(\onlinecite{Martin}). A practical scheme to control the sloshing
instability is discussed in the paper.

To summarize, we improve the numerical efficiency of fictitious
dynamical methods for electrons in several ways. Firstly, we replace
steepest descent dynamics by a more efficient damped second order
dynamics to minimize the total energy. Secondly, by preconditioning the
fictitious electronic masses we increase the integration time step
for total energy minimization and for simulation of the
adiabatic ionic dynamics. Thirdly, we show that in the context of
fictitious dynamical methods the so called `charge sloshing' problem, which
is expected to arise for large systems, is less serious than expected.
We support our theoretical analysis with detailed numerical tests
on several systems involving Si and C atoms.

The paper is organized as follows. In Sec. II we discuss first-order
SD dynamics and second-order conservative MD dynamics for the
electronic degrees of freedom. In Sec. III we introduce a damped
second-order dynamics which is substantially more efficient than SD and is
competitive with the best conjugate gradient schemes for
electron minimization.
In Sec. IV we discuss large wavevector instabilities and the `charge sloshing'
problem.
In Sec. V we discuss the preconditioning of large wavevector components.
In Sec. VI we present some details of the numerical
implementation. Finally, in Sec. VII we present the results of realistic
numerical tests on silicon and carbon systems.
Sec. VIII is devoted to our conclusions.

\section{Fictitious dynamics for the electrons}

Dynamical methods for minimizing the
electronic total energy and for simulating the
adiabatic motion of the atoms are based on a
fictitious dynamics of the electronic degrees of freedom.
Within these approaches the forces acting on the electronic degrees of
freedom are
derived from the total electronic energy $E[\{\psi\}]$ in the DFT-LDA
form:
\be E[\{\psi\}]=E_{kin}[\{\psi\}]+E_{ext}[\rho]+E_{h}[\rho]+
E_{xc}[\rho],\ee
where $E_{kin}$, $E_{ext}$, $E_{h}$, and $E_{xc}$
denote kinetic, external potential, Hartree and exchange-correlation
energy, respectively\cite{DFT}. The local density approximation
is adopted for the latter\cite{DFT}.
The electronic charge density $\rho$ is given by:
\be \rho ({\bf r})=2<\Psi_i|{\bf r}><{\bf r}|\Psi_i>,\ee
where the occupied orbitals $|\Psi_i>$ are orthonormal.
The factor of 2 accounts for the occupation numbers,
which here and in the following are supposed to be all equal to 2.
Summation over repeated indices is understood.

In order to ensure the orthonormalization of the
electronic orbitals during a dynamical evolution it is convenient to add
appropriate forces of constraints. These do
not perform work on the electronic system, and can be conveniently
calculated in terms of Lagrangian multipliers.
The corresponding equations of motion for the first order
dynamics are:
\be \mu |\dot \psi_i>=-\frac{1}{2}
\frac{\delta E[\{\psi\}]}{\delta \psi_i}
+\Lambda_{ij}|\psi_j>,\label{O1}\ee
and those for the  second order dynamics are:
\be \mu |\ddot \psi_i>=- \frac{1}{2}
\frac{\delta E[\{\psi\}]}{\delta \psi_i}
+\Lambda_{ij}|\psi_j>.\label{O2}\ee
where we have assumed that the wavefunctions $|\psi>$ are real. Here
and in the following the indices $i$ and $j$ run over the occupied states
only. The
symmetric matrix $\Lambda_{ij}$ of the Lagrange multipliers enforces
the orthonormality condition, i.e.  $<\psi_i|\psi_j>=\delta_{ij}$.
The derivatives of $E[\{\psi\}]$ with
respect to the $|\psi_i>$ define the Kohn-Sham Hamiltonian:
\be \frac{1}{2}\frac{\delta E[\{\psi\}]}{\delta\psi_i}=
\hat 2 H_{KS}|\psi_i>.\ee
The parameter $\mu$ is a fictitious electronic mass. It is used to
tune the speed of the electronic dynamics and does not describe any
other physical property. When the ions are held fixed, this mass can
be included in the definition of the time step and it is
irrelevant. However when we allow the ions to move, the ratio between
$\mu$ and the physical ionic masses is important since it defines the
relative speed of the ionic and of the fictitious electronic motion.

Now let us suppose that the wavefunctions are close to the minimum of
the energy $E[\{\psi^{(0)}\}]$,
\be | \psi_i>=|\psi^{(0)}_i>+|\delta \psi_i>,\ee
where $|\psi^{(0)}_i>$ are the wavefunctions at the minimum and
$|\delta \psi_i>$ are the corresponding deviations. We notice that
$|\delta \psi_i>$ have to
fulfill the orthonormality condition to linear order, i.e.
$<\psi^{(0)}_i|\delta \psi_j>=0$. To linear order in
$|\delta \psi_i>$, the Lagrange multipliers $\Lambda_{ij}$ are the same
for  first and second order dynamics, and  are given by
\be \Lambda_{ij}=2<\psi_i|H_{KS}|\psi_j>\label{eq:lambda}.\ee
Thus, by retaining only the
terms up to linear order in $|\delta \psi_i>$ in the equations of
motion  (\ref{O1}) and (\ref{O2}) we obtain:
\be |\delta \dot \psi_i>=-\hat \om_{ij}|\delta \psi_j>
\label{linO1}\ee
and
\be |\delta \ddot \psi_i>=-\hat \om_{ij}|\delta \psi_j>,
\label{linO2}\ee
respectively. Here $\hat \om_{ij}$ is a linear operator, which acts on
the single-particle Hilbert space and which has the {\em same} form for {\em
both} first- and second-order dynamics.
In the following we will use the notation $\hat \bom$ to
indicate a matrix of operators having for elements the $\hat \om_{ij}$.
Notice that $\hat \bom$ is a positive definite linear
operator since Eqs. (\ref{linO1}) and (\ref{linO2})  result from a quadratic
expansion of $E[\{\psi\}]$ in $|\delta \psi_i>$
about the minimum  $E[\{\psi_0\}]$, i.e.:
\be
E[\{\psi\}] - E[\{\psi_0\}]=\mu<\delta \psi_i|\hat \om_{ij}
|\delta \psi_j>+O(\delta \psi^3).\label{e2o}
\ee
The equations of motion (\ref{linO1}) and (\ref{linO2}) can be formally
integrated yielding:
\be |\delta \psi_i(t)> = \left(exp(-\hat {\bom}t)\right)_{ij}|
\delta \psi_j(0)>\label{sO1}\ee
and
\be |\delta \psi_i(t)> = (cos{\sqrt{\hat {\bom}}t })_{ij}
|\delta \psi_j(0)>+({\hat {\bom}}^{-1/2}
sin{\sqrt{\hat {\bom}}t})_{ij}|\delta \dot \psi_j(0)>\label{sO2},\ee
respectively. In the case of first order evolution the
wavefunctions decay exponentially towards the  minimum
$E[{\psi}_0]$, while in the second order evolution they
perform small oscillations around it. These motions take place with
characteristic decay rates and frequencies which are equal to the
eigenvalues $ \om_{\alpha}$ of the operator $\hat\bom$ and to the
square root $\sqrt{\om_{\alpha}}$ of these eigenvalues
for first and second order dynamics, respectively.

In numerical implementations the electronic states
are expanded on a finite basis set, so that only a finite number of
eigenfrequencies and eigenmodes occur.
Let $\om_{min}$ and $\om_{max}$ be the minimum and
maximum eigenvalues of $\hat \bom$. The maximum allowed
time-step for numerical integration is proportional to the
smallest period of the system, i.e. to $1/\om_{max}$ or to
$1/ \sqrt{\om_{max}}$ for first and second order dynamics, respectively.

In the case of first order dynamics, the minimum eigenvalue dominates
the long-time behavior of the decay to the ground-state, so
that a rough estimate of the convergence time is given by
$1/\om_{min}$. Recalling that the size of the time step is
proportional to $1/\om_{max}$, one finds that the number $n_{O1}$ of
integration steps needed to converge satisfies the
condition:
\be n_{O1} \propto \om_{max}/\om_{min}.\label{nO1}\ee

In the case of second order dynamics we usually start
a simulation from an electronic
configuration close to the minimum of the electronic energy. Then
if the ionic and the
electronic frequencies are well decoupled \cite{Car,lh},
the electrons remain adiabatically close to the
instantaneous energy minimum during the ionic evolution.
Let $\omega_{ion}$ be a typical ionic frequency.
The adiabatic
condition requires it to be much smaller than the
minimum electronic frequency i.e.
\be   \omega_{ion}<<\sqrt{\om_{min}}.\label{decouple}\ee
A meaningful measure of the simulation's workload is given by the
number of time steps necessary to integrate a full ionic oscillation.
Thus, recalling that the time step is inversely proportional to
$\sqrt{\om_{max}}$, we find that the number $n_{O2}$
of steps necessary to integrate a typical ionic oscillation satisfies
the following condition:
\be n_{O2}\propto \sqrt{\om_{max}/\om_{min}}.\label{nO2}\ee

\section{Damped second order dynamics for minimization}

In the previous section we showed that the number of iterations necessary
to minimize the electronic energy within steepest descent dynamics is
proportional to $\om_{max}/\om_{min}$. In this section we present an
improved {\em minimization} dynamics in which the typical number of
iterations is instead proportional to $\sqrt{\om_{max}/\om_{min}}$,
i.e. a number significantly smaller than  $\om_{max}/\om_{min}$.
We attain this goal by inserting in Eq. (\ref{O2})
a damping term as follows:
\be \mu |\ddot \psi_i>=-\frac{1}{2}\frac{\delta E[\{\psi\}]}{\delta
\psi_i}-2\gamma\mu |\dot \psi_i>
+\Lambda_{ij}|\psi_j>\label{DO2}\ee
This equation defines a damped second order dynamics. As in the previous
section we study the resulting motion close to the energy
minimum. We find that the deviations of the wavefunctions from the minimum
are subject to damped oscillations given by:
\be |\delta \psi_i(t)>=exp(i\hat\bo_{+} t)_{ij}|\psi_j^{(+)}>+
exp(i\hat\bo_{-} t)_{ij}|\psi_j^{(-)}>\label{sDO2}\ee
Here $|\psi_j^{(+)}>$ and $|\psi_j^{(-)}>$ are determined by the initial
conditions, and \be\hat\bo_{\pm}=i\gamma \ide \pm\sqrt{\hat
\bom-\gamma^2 \ide}\ee The real part of $\hat\bo_{\pm}$ gives the
frequencies of the oscillatory motion, while its imaginary part
gives the decay rate to the minimum. In order to maximize the
rate of convergence, we must use the maximum value of $\gamma$
for which the argument of the square root remains positive.
This optimal value of $\gamma$ is given by
\be \gamma_{opt}=\sqrt{\om_{min}},\label{gammopt}\ee
since this value corresponds to critical damping of the smallest eigenvalue of
$\hat \bom$. In this case the imaginary part of all the eigenvalues of
$\hat\bo_{\pm}$ is equal to $\gamma_{opt}$ and the time of
convergence to the minimum is of the order of $1/\sqrt{\om_{min}}$.
The integration time step is related to
the maximum norm of the eigenvalues of $\hat\bo_{\pm}$, which is equal
to $\sqrt{\om_{max}}$.
Thus, the number of
integration steps necessary for minimization is given by:
\be n_{DO2}\propto \sqrt{\frac{\om_{max}}{\om_{min}}}\propto
\sqrt{n_{O1}}.\label{nDO2}\ee
{}From this formula we see that a relevant gain of efficiency
is obtained when using damped dynamics instead than steepest-descent
dynamics to minimize the electronic energy.
The gain is particularly important when a large
number of iterations is needed to converge to the ground-state, which is
typically the case of metallic systems.

\section{ Second order expansion of the LDA energy functional}

In this section we compute explicitly the
eigenvalues of the operator $\bom$.
For this purpose we consider the  expansion
of the energy functional around its minimum $\Psi^{(0)}$ up to second order in
$\delta \Psi$. This is given by:
\be E[\{\Psi\}]-E[\{\Psi^{(0)}\}] = 2<\delta \Psi_i|H_{KS}|\delta \Psi_i>
-2<\delta \Psi_i|\delta
\Psi_j><\Psi^{(0)}_i|H_{KS}|\Psi^{(0)}_j>\label{e2nd}\ee
$$+\int d{\bf r}\int d{\bf r}' \delta \rho ({\bf r})
\left[ \frac{\delta^2 E_{h}}{\delta \rho ({\bf r}) \delta \rho ({\bf r}')}
+\frac{\delta^2 E_{xc}}{\delta \rho ({\bf r}) \delta \rho ({\bf r}')}\right]
\delta \rho ({\bf r}')+O(\{\delta \Psi^3\}).$$
The second term on the r.h.s. of this equation
comes from the Lagrange multipliers (see
Eq. (\ref{eq:lambda}) ),
and
\be  \delta \rho ({\bf r})=4<\Psi^{(0)}_i|{\bf r}><{\bf r}|\delta \Psi_i>
\label{dcharge}\ee
gives the variation of the electronic density to first order in $\{\delta
\Psi\}$. We recall that $\{\Psi\}$ and $\{\delta\Psi\}$ are supposed
to be real.

\subsection{Non self-consistent case}

If we neglect the last two terms in
Eq. (\ref{e2nd}), i.e. the terms corresponding to
variations of the Hartree and exchange-correlation
potentials, we recover the expansion of the total
energy appropriate to a non-selfconsistent Hamiltonian.
Then we can expand the $\Psi^{(0)}$ and $\delta \Psi$
in terms
of the the real eigenvectors $| \chi^0_l >$ of $H_{KS}$, which have eigenvalues
$\varepsilon_l$.
Since the total energy is invariant under unitary
transformations in the subspace of occupied states,
we can suppose without loss of generality:
$$
| \Psi^{(0)}_i> =| \chi^0_i>,
$$
\begin{equation}
| \delta \Psi_i> =  \sum_{k}  c^i_k | \chi^0_k >,\label{eigexp}
\end{equation}
where $ c^i_k$ are real coefficients.
Here and in the following the indices $i$ and $k$ refer to
occupied and unoccupied states, respectively.
Hence, as shown in Ref.~(\onlinecite{Pastore}), we obtain for $\delta E_{nsc}$,
i.e.
the second-order variation of
the energy in which the selfconsistency of the
potential is not taken in account:
\begin{equation}
\delta E_{nsc} = 2\sum_{ik} (c^i_k)^2 (\varepsilon_k -
\varepsilon_i).\label{Ensc}
\end{equation}
where $\varepsilon_i$ and $\varepsilon_k$
are respectively the occupied and the unoccupied eigenvalues of
$ H_{KS}$. By comparing Eq. (\ref{Ensc}) with Eq. (\ref{e2o})
we see that
the eigenvalues of $\hat \bom$ are given by \be
\om_{(i,k)}=2\frac{\varepsilon_k-\varepsilon_i}{\mu}\label{kappaik}\ee
and the
lowest eigenvalue of $\hat \bom$ is given by $\om_{min}=2E_{gap}/\mu$,
in terms of the energy gap $E_{gap}$ separating the lowest unoccupied
from the highest occupied electronic level.

In the case of an insulator, the energy gap has a finite positive value
which, above a certain size, is independent of the simulation cell.
In the case of a metal instead, the energy gap is still finite and
positive for a finite
sized system but it is no longer independent of the simulation cell. In fact
the energy gap and $\om_{min}$ tend to zero for a cell size going to infinity.
However, many properties of interest do not require an infinite
energy resolution for the states around
the Fermi energy. Typically a small but finite energy resolution $E_{err}$
is sufficient.
$E_{err}$ does not depend on the size of the system, and $K_{err}=2E_{err}/\mu$
replaces $K_{min}$ in  Eq.s (\ref{nO1}),(\ref{nDO2}) and (\ref{gammopt})
to estimate  the convergence rates $n_{O1}$,
$n_{DO2}$ and the optimal damping parameter $\gamma_{opt}$.
Since $E_{err}$ is much smaller than a typical energy gap of an insulator,
the number of iterations needed for ground state convergence is
much larger in metals than in insulators.
For the same reason, a perfectly adiabatic
separation between ionic and electronic motions is not
possible for metals. However, as shown in Ref.~(\onlinecite{Bloechl}) a
satisfactory
solution to this
problem, in the context of Car-Parrinello simulations, can be obtained
by using two Nose' thermostats to control separately
the respective temperatures of the ions and of the electrons.

When expanding the wavefunctions in terms of plane-waves,
we can define an effective cut-off energy $E_{cut}$ given (in a.u.) by
$q_{max}^2/2$, where $q_{max}$ is the largest wavevector in the
basis set. The band
of empty states is usually much larger than the band of occupied states.
Thus when in Eq. (\ref{kappaik}) the index $k$ refers to the highest
unoccupied states, the
eigenvalues $K_{(i,k)}$ of $\hat \bom$ have a negligible dependence on
the occupied state index $i$.
Furthermore, since the highest unoccupied
states are free-particle like, the energy difference
$\varepsilon_k-\varepsilon_i$ is dominated by the kinetic energy of
the state $k$, i.e.
\be \varepsilon_k-\varepsilon_i\sim q^2/2,\label{epsdiff}\ee
where $q$ is the wavevector associated to the state $k$. The maximum
eigenvalue of $\hat \bom$ is therefore approximately given by
$\om_{max}\simeq 2(q_{max}^2/2)/\mu = 2E_{cut}/\mu$.
It is this eigenvalue that limits the maximum allowed time step for numerical
integration in a non self-consistent case: the numerical integration becomes
unstable and the time step has to go to zero when $ E_{cut}$
goes to infinity.

\subsection{Self-consistent case: charge sloshing}

We now consider
the terms of Eq. (\ref{e2nd}) that we neglected in the previous subsection
in order to see if they affect the maximum
eigenvalue of $\hat \bom$. In this case $\hat \bom$ is not diagonal in the
representation of the $c^i_k$ and, for an arbitrary system, it is not
possible to
diagonalize it analytically. Thus we need some simplifying assumptions.
Let us consider a crystal of given periodicity and
use a supercell containing an arbitrary number of replicas of the
crystal unit cell. In this case the $\{\Psi_0\}$ are linear combinations of
Bloch-functions with
the crystal periodicity whereas the fluctuations $\{\delta \Psi\}$ may have all
the wavelengths compatible with the supercell. In other words we are
restricting the periodicity of the unperturbed state but not the
periodicity of the
fluctuations. Based on the above simplifying assumption we find that
charge sloshing affects differently metallic and non-metallic systems.
A numerical example presented in Sec. VII suggests that this result should
hold also for non-periodic systems.

In order to find out whether the maximum eigenvalue of $\hat \bom$
diverges when the supercell size tends to infinity,
we restrict our analysis to the Hartree term since the LDA exchange-correlation
energy is well behaved and typically
has a negligible effect compared to the kinetic
energy on the maximum eigenvalue of $\hat \bom$.
The second order variation of the Hartree energy is given by:
\be
\delta E_H=\int d{\bf r}\int d{\bf r}' \delta \rho ({\bf r})
{1\over |{\bf r} - {\bf r}'| }
\delta \rho ({\bf r}')=\sum_{\bf G}\sum_{\bf p}{4\pi\over \Omega |{\bf p}+
{\bf G}|^2}
|\delta\rho({\bf p}+{\bf G})|^2.\label{dEH}
\ee
Here {\bf p} is a vector belonging to the first Brillouin Zone of the crystal,
{\bf G} is a vector of the reciprocal lattice of the crystal and
the sums extend over all the non-zero wavevectors
${\bf p} + {\bf G} = {\bf q} - {\bf q}'$ where ${\bf q}$ and ${\bf q}'$ are
two generic plane-waves of the basis set used to represent the electron
wavefunctions in the supercell of volume $\Omega$. $\delta\rho({\bf p} +
{\bf G})$
is the Fourier Transform (FT) of $\delta \rho ({\bf r})$.

When a linear dimension $L$ of the supercell becomes very large,
${({\bf p} + {\bf G})}_{min}=({\bf p})_{min}$,
i.e. the smallest nonzero ${\bf q} - {\bf q}'$ vector,
tends to zero like $1/L$. If, correspondingly, the maximum eigenvalue of
$\hat \bom$ diverges, we have the so-called `charge sloshing' scenario.
To study the effect of $ {\bf p}_{min}$ on the maximum eigenvalue of $\hat
\bom$
we consider only the terms
with ${\bf G}=0$ in Eq. (\ref{dEH}). Then
since $\delta \rho ({\bf r})$ is real,
$|\delta\rho({\bf p})|=|\delta\rho(-{\bf p})|$ and $\delta E_H({\bf G}=0)$
can be written as:
\be
\delta E_H({\bf G}=0)=
\sum_{ {\bf p},p_x>0}{8\pi\over \Omega  p^2 }
|\delta\rho({\bf p})|^2,\label{dEH0}
\ee
where
\be
\delta \rho({\bf p}) =
\sum_{ik}4(< \chi^0_i |cos({\bf p r})| \chi^0_k >c^i_k+i
< \chi^0_i |sin({\bf p r})| \chi^0_k >c^i_k)
\ee
Notice that Eq. (\ref{dEH0}) is a quadratic form in terms of the $c^i_k$.
The real coefficients
$c^i_k$ can be considered as the components of a real vector $\bf |c>$.
Similarly
we can introduce two real vectors $\bf|A(p)>$ and $\bf|B(p)>$ whose components,
labelled by the composite index $(ik)$, are given by
$< \chi^0_i |cos({\bf p r})| \chi^0_k >$ and
by $< \chi^0_i |sin({\bf p r})| \chi^0_k >$,
respectively. In this notation $\delta \rho({\bf p}) = 4({\bf <c|A(p)>}+
i{\bf <c|B(p)>})$ and Eq. (\ref{dEH0}) becomes:
\be
\delta E_H({\bf G}=0)=
\sum_{{\bf p},p_x>0}{8\pi\over \Omega  p^2
}({\bf <c|A(p)><A(p)|c>}+{\bf <c|B(p)><B(p)|c>}).\label{dEH0bis}
\ee
Using the fact that the $\chi^0_l$ are eigenstates of a periodic crystal,
it is easy to show\cite{pp'} that
the vectors $\bf|A(p)>$ and $\bf|B(p)>$ constitute an orthogonal set:
\be
\begin{array}{rcl}
{\bf<A(p)|A(p}'{\bf)> }&=&\delta_{{\bf p},{\bf p}'}{\Omega\over 2}
S({\bf p})\nonumber\\
{\bf<B(p)|B(p}'{\bf)> }&=&\delta_{{\bf p},{\bf p}'}{\Omega\over 2} S({\bf p})\\
{\bf<B(p)|A(p}'{\bf)> }&=&0\nonumber
\end{array}
\ee
where $p_x,p'_x>0$ and the `static structure factor' $S({\bf p})$
is defined by:
\be
S({\bf p})={1\over \Omega}\sum_{ik}< \chi^0_i |e^{-i{\bf p r}}| \chi^0_k >
< \chi^0_k |e^{+i{\bf p r}}| \chi^0_i>.\label{S(p)}
\ee
Hence $\bf|A(p)>$ and $\bf|B(p)>$ are the vectors that diagonalize the
quadratic form in Eq. (\ref{dEH0bis}).
The corresponding eigenvalues of $\hat \bom $ are given by:
\be
K_{\bf A(p)} = K_{\bf B(p)}= {128\pi\over \mu \Omega p^2}{\bf<A(p)|A(p)> }=
{128\pi\over \mu \Omega p^2}{\bf<B(p)|B(p)> }=
{64\pi\over \mu} {S({\bf p})\over p^2}
\ee
Therefore, when a linear dimension $L$ of the supercell tends to infinity and,
correspondingly,  $ {\bf p}_{min}$ goes to zero,
$K_{ {\bf A(p}_{min}{\bf)}}$ and
$K_{ {\bf B(p}_{min}{\bf)}}$
do not diverge if $S({\bf p})$ is of order $O(p^2)$.

We now consider a jellium model as a representative metallic system.
In this case the $\chi^0_i$ are plane waves and one finds\cite{je}:
$ S({\bf p}) =p[ 1-(p/p_F)^2/12]p_F^2/8\pi^2$, where $p_F$ is the
Fermi momentum, and $p<2p_F$.
As a consequence for $L$ going to infinity,   $K_{ {\bf A(p}_{min}{\bf)}}$ and
$K_{ {\bf B(p}_{min}{\bf)}}$ diverge as $L$ and the
time step for numerical integration has to be reduced accordingly: this is
a charge sloshing situation.

When the system is a periodic insulator one finds instead
that $S({\bf p})$ goes to zero as $p^2$ (see Appendix).
As a consequence for $L$ going to infinity,
$K_{ {\bf A(p}_{min}{\bf)}}$ and
$K_{ {\bf B(p}_{min}{\bf)}}$ tend to a constant and
the time step for numerical integration is independent of $L$:
charge sloshing is absent here.

We stress that the above conclusions apply only if we consider small
fluctuations around the ground-state: this is the typical case of
ab-initio molecular dynamics simulations of the ionic motion.
However, in the initial steps of an electronic minimization
procedure, the wavefunctions
may be far from the ground-state. In this case it is possible to observe
sloshing effects also in periodic systems having an insulating ground-state.

Since charge-sloshing is a consequence of the singularity of the Coulomb
potential at small $\bf p$, a simple way of eliminating charge-sloshing
instabilities
consists in replacing the Coulomb potential $4\pi/p^2$
with a Yukawa potential $4\pi/(p^2+\alpha^2)$, where $2\pi/\alpha$ is a typical
decay length of the order of the system size $L_{min}$ that corresponds to the
onset of the sloshing instability. In the case of an insulator
we can use this technique to stabilize the numerical integration during
the initial steps of an electronic minimization run. Then when we are
sufficiently close to the ground-state we can set $\alpha = 0$ and converge to
the exact ground-state. We will show with a numerical example in a subsequent
section that this technique allows us to converge to the exact ground-state
of a disordered insulating system with a number of steps independent
of the system size. In the case of a metal it is not possible to set $\alpha$
equal to zero not even in the proximity of the ground-state. However we
notice that $L_{min}$ is usually much larger than the typical screening length
of a metal. The results of a numerical simulation for a large but finite
metallic system should not change appreciably if the Coulomb potential
is replaced by a Yukawa potential that is equal to the Coulomb potential
for distances smaller than $L_{min}$.

\section{Preconditioning the equations of motion}

The numerical efficiency of all the fictitious
dynamical methods previously introduced can be improved
by preconditioning the dynamics in order to reduce
the ratio $\om_{max}/\om_{min}$. This can be achieved by replacing
the constant fictitious mass parameter $\mu$ in Eqs. (\ref{O1},\ref{O2},
\ref{DO2}).
with an arbitrary positive definite operator $\hat \mu$.
The resulting increased arbitrariness in the choice of $\hat \mu$
can be exploited to
compress the highest frequency components of the spectrum of the fictitious
electron dynamics. Recalling that these are due basically to the high energy
unoccupied states which are free-particle like
(see Eq. (\ref{epsdiff})), we choose an operator $\hat \mu$ which is diagonal
in
$q$-space with eigenvalues $\mu(q)$ given by:
\be \begin{array}{ll}\mu(q)=\mu_0&{\rm if}~~ \frac{1}{2}q^2<E_p\\
\mu(q)=\mu_0\frac{q^2}{2E_p}&{\rm if}~~
\frac{1}{2}q^2>E_p\end{array}\ee
Below a certain cutoff energy $E_p$, it is worth considering a constant mass
$\mu_0$,
because the low energy eigenstates have a relevant potential energy
contribution and are not free-particle like. The preconditioning
cut-off $E_p$ therefore represents the threshold above which the
states are dominated by the kinetic energy.

It is easy to show that the solutions of the preconditioned equations
of motions for small displacements, are still given by Eqs.
(\ref{sO1},\ref{sO2},\ref{sDO2}) if the operator $\hat \bom$ is
replaced by the operator $\hat{\bar \bom}$ characteristic of the
preconditioned dynamics. All the relations  (\ref{nO1}), (\ref{nO2}),
(\ref{nDO2}) found for first and second order dynamics with and
without damping hold therefore also in the case of the preconditioned
dynamics but, in the latter case,
$\om_{max}$ and $\om_{min}$ have to be replaced by the maximum
and minimum eigenvalues of $\hat {\bar \bom}$, i.e. by $\bar
\om_{max}$ and $\bar \om_{min}$.

The preconditioning cut-off $E_p$ that minimizes the ratio $\bar
\om_{max}/\bar \om_{min}$ is called the optimal preconditioning cutoff.
It depends strongly on the
atomic species, i.e. on the pseudopotentials and on
the plane-wave cutoff that are used in the calculation.
It depends only negligibly on the physical environment.
Thus, for a given atomic species, it is possible to find the optimal
preconditioning cutoff by performing calculations on
a simple reference
system. We present a typical example in Sec. VII.

\section{Numerical implementation}

In our numerical implementation we adopt the standard procedures
described in Refs. (\onlinecite{lh,review}) to integrate the equations of
motion
for first and second order dynamics.
In the case of damped second order dynamics we follow the
procedure introduced in Ref.~(\onlinecite{Bloechl}) to integrate Car-Parrinello
dynamics in presence of a Nos\'e-Hoover thermostat. We obtain for
first, second order and damped dynamics, respectively:
\be |\psi_i(t+\Delta)>=|\psi_i(t)>-2
\hat \mu^{-1}\hat H_{KS}|\psi_i> \Delta
+X_{ij}\hat \mu^{-1}|\psi_j(t)>\label{iO1}\ee
\be |\psi_i(t+\Delta)>=-|\psi_i(t-\Delta)>+2|\psi_i(t)>-2
\hat \mu^{-1}\hat H_{KS}|\psi_i> \Delta^2
+X_{ij}\hat \mu^{-1}|\psi_j(t)>\label{iO2}\ee
$$  |\psi_i(t+\Delta)>=|\psi_i(t-\Delta)>+
\phantom{\left(2|\psi_i(t)>-|\psi_i(t-\Delta)>-2
\hat \mu^{-1}\hat H_{KS}|\psi_i> \frac{\Delta^2}{2}\right)}$$
\be +\left(|\psi_i(t)>-|\psi_i(t-\Delta)>-
\hat \mu^{-1}\hat H_{KS}|\psi_i> \frac{\Delta^2}{2}
\right) \frac{2}{1+\gamma\Delta}
+X_{ij}\hat \mu^{-1}|\psi_j(t)>\label{iDO2}\ee
where $\Delta$ is the integration time step and $X_{ij}$ is a
symmetric matrix equal to $\Lambda_{ij}\Delta$ for   first order
dynamics, equal to $\Lambda_{ij}\Delta^2$ for  conservative second
order dynamics, and equal to  $\Lambda_{ij}\Delta^2/ (1+\gamma\Delta)$
for  damped second order dynamics. The matrix $\bf X$ is found by
imposing the orthonormality of the wavefunctions at time
$t+\Delta$:
\be <\psi_i(t+\Delta)|\psi_j(t+\Delta)>=\delta_{ij}\label{orto1}\ee
We notice that the inversion of the mass operator $\hat \mu$ is
straightforward in $q$ space where it is diagonal. For the calculation
of  $ X_{ij}$ we define the wavefunctions $|\bar \psi_i(t+\Delta)>$
as the  r.h.s. of the Eq. (\ref{iO1},\ref{iO2},\ref{iDO2}) without the
orthonormalization terms $X_{ij}\hat \mu^{-1}|\psi_j(t)>$. Then  Eq.
(\ref{orto1}) becomes:
\be  {\bf XMX}^{\dagger}+{\bf BX}^{\dagger}+{\bf X B}^{\dagger}=
{\bf 1-A}\label{lambda}\ee
where  the matrices $\bf M, B, A$ are given respectively by:
\be M_{ij}=<\psi_i(t)|\hat\mu^{-2}|\psi_j(t)>\ee
\be B_{ij}=<\bar \psi_i(t+\Delta)|\hat\mu^{-1}|\psi_j(t)>\ee
\be A_{ij}=<\bar \psi_i(t+\Delta)|\bar \psi_j(t+\Delta)>.\ee
The scalar products are easily evaluated in $q$-space where the mass
operator $\hat \mu$ is diagonal.

Eq. (\ref{lambda})
is formally identical to the
matrix equation that expresses the orthonormality condition
for Car-Parrinello dynamics when using Vanderbilt's
ultrasoft pseudopotentials \cite{Laasonen}. It can be solved as
described in Ref.~(\onlinecite{Laasonen}). The
matrix $\bf B$ can be conveniently split into a symmetric part $\bf
B_s$ and an antisymmetric  part $\bf B_a$. The antisymmetric part $\bf
B_a$ is first order in $\Delta$, while $\bf X$ and $\bf 1-A$ are
first (second) order in $\Delta$ for first (second) order dynamics.
Using these properties, we can solve Eq. (\ref{lambda})
iteratively in terms of increasing powers of $\Delta$ ($\Delta^2$):
\be {\bf B_sX}^{(n+1)}+{\bf X}^{(n+1)} {\bf B_s}=
{\bf 1-A-X}^{(n)}{\bf  M X}^{(n)}-{\bf B_aX}^{(n)}+
{\bf X}^{(n)}{\bf B_a}
\label{itlambda}\ee
Here ${\bf X}^{(0)}$ is the solution of the equation:
\be {\bf B_s X}^{(0)}+{\bf X}^{(0)} {\bf B_s=1-A}\label{x0}\ee
and the l.h.s of the Eq. (\ref{itlambda},\ref{x0}) is inverted
after transforming to the basis where $\bf B_s$ is diagonal.

An alternative approach based on unconstrained total energy functional which
avoids explicit orthonormalization has
been recently proposed in Refs. (\onlinecite{Mauri,Mauri2}). The electronic
mass
preconditioning scheme discussed in the present paper
can be easily applied to the unconstrained energy functional method
without any overload.

In some applications of fictitious dynamical methods for electrons the
orthonormalization of the electronic wavefunctions can be achieved via a
Gram-Schmidt procedure. We stress that this approach is not justified in
connection with the mass preconditioning scheme described above.
Indeed if the Gram-Schmidt orthogonalization procedure is used within
preconditioned steepest-descent dynamics, one is not guaranteed that the energy
will decrease at any integration step for a sufficiently small time step.
The origin of this instability is related to the
non-holonomic character of the constraints imposed via a Gram-Schmidt
procedure. We found that this instability is rather severe in practical
numerical applications, where it spoils all the efficiency gains of the mass
preconditioning scheme.

\section{Numerical Results}

We tested the different dynamical schemes described above on various
physical systems within a DFT-LDA formulation.
In particular we considered Si and C systems.
We used pseudopotentials of the Bachelet-Hamann-Schl\"uter
type \cite{Bachelet}, with $s$ and $p$ non-locality in the
Kleinmann-Bylander form  \cite{Kleinmann}.
The cut-off for the plane-wave expansion of the electronic orbitals was
12 Ry for silicon, and 35 Ry for carbon.
We carried out all the
calculations at the $\Gamma$ point of the Brillouin zone only.
Moreover, in order to compare the
dynamical
schemes with conjugate gradient minimization, we used a
tight-binding energy functional for carbon\cite{Wang}.

\subsection{preconditioning}

We start by presenting the results obtained with preconditioning. In
order to determine the optimal preconditioning cut-off $E_p$ we had to
minimize the ratio $\rat$. We measured $ \bar \om_{max}$ and $ \bar
\om_{min}$ within first order and second order dynamics, by giving a
small displacement to  the system from its energy minimum. In the case
of first order dynamics, the numerical integration of Eq. (\ref{iO1}),
becomes unstable and results in an exponential increase of the energy,
when $\Delta>2/\bar \om_{max}$. Therefore the maximum allowed
integration time step provides an accurate way of estimating $ \bar
\om_{max}$. $\bar \om_{min}$, i.e. the lowest eigenvalue of $\hat
{\bar \bom}$, gives instead the slowest rate of decay of the energy.
This rate is conveniently sampled at large times, i.e. when only the
slowest exponential is left in the decay.

We report in Fig. (1) the ratio $\bar\om_{max}/\bar\om_{min}$ as a
function of $E_p$ for a Si$_3$ molecule. We notice that for the highest
$E_p$ values the ratio decreases linearly with decreasing $E_p$. The
behavior of the ratio $\bar\om_{max}/\bar\om_{min}$ in this range is
explained by the following considerations. First, the minimum
frequency is unchanged, since it is related to the lowest excited
state which has small components at high $q$. Second, all the excited
modes at energies higher than $E_p$ are compressed to the same maximum
frequency $2E_p/\mu_0$, as long as they are kinetically dominated.
Instead, in the range of low preconditioning cut-offs $E_p$, the
minimum frequency $\bar \om_{min}$ decreases and the highest excited
modes become less efficiently compressed. Thus a minimum value of the
ratio $\bar \om_{max}/\bar \om_{min}$ is found, as we can see in
Fig. (1). This minimum occurs at $E_p$=1 Ry. The corresponding reduction
of the ratio  $\bar K_{max}/\bar K_{min}$ is of a factor of 5 compared
to the non-preconditioned case. We obtained very similar results for a
sample of crystalline silicon in the diamond structure,
where the optimal $E_p$ was also close to 1 Ry.

In the case of second order dynamics $ \sqrt{\bar \om_{max}}$ and $
\sqrt{\bar \om_{min}}$ can be found as the maximum and minimum
frequencies of the power spectrum of the fictitious electronic
dynamics. This is easily evaluated by computing the velocity
autocorrelation function corresponding to the wavefunction dynamics.
The power spectrum of the velocity autocorrelation function
corresponding to the electronic fictitious dynamics is given in Fig. (2)
for the case of the Si$_3$ molecule. In particular we show results
obtained with optimal preconditioning ($E_p=1$ Ry) and
unpreconditioned dynamics. In calculating the spectra we chose the
value of $\mu_0$ in such a way that the lowest frequency $ \sqrt{\bar
\om_{min}}$ of the preconditioned dynamics coincided with $
\sqrt{\om_{min}}$, i.e. the lowest frequency of the dynamics without
preconditioning. This is achieved by setting $\mu_0$=260 a.u. when the
mass associated to the dynamics without preconditioning is $\mu$=300
a.u.. The significant compression of the high frequency modes
resulting from preconditioning is clearly evident in Fig. (2).

For a
sample of crystalline carbon in the diamond structure we found an optimal value
of $E_p$
equal to 2.7 Ry. This reduced by a factor of 9 the ratio $\rat$
compared to the unpreconditioned case. In this case in order to make
the lowest frequencies $ \sqrt{\bar \om_{min}}$ and $
\sqrt{\om_{min}}$ to coincide, the mass $\mu$ of the unpreconditioned
dynamics had to be rescaled by a factor of 0.93 in order to obtain the
mass $\mu_0$ of the preconditioned dynamics.

We notice that in a different context the authors of Ref.~(\onlinecite{Teter})
proposed to use a preconditioning cutoff $E_p$ equal to the
expectation value of the kinetic energy  divided by the number of
electrons. In the cases discussed above this corresponds to a value of
$E_p=$ 0.8 Ry and $E_p=$2 Ry for Si and C, respectively. These values
are close to the optimal values of $E_p$.

\subsection{Ionic Molecular Dynamics}

In order to test the effect of preconditioning on ab-initio molecular
dynamics simulations of the ionic motion, we considered the coupled
set of equations given by Eq. (\ref{O2}) for the electronic degrees of
freedom and by
\be M_i \ddot {\bf R}_i=-
\frac{\delta E[\{\psi\},{\bf R}]}{\delta {\bf R}_i}\label{ionsdin}\ee
for the ionic coordinates ${\bf R}_i$. Here $ M_i$ are the physical
ionic masses and the mass $\mu$ in Eq. (\ref{O2}) has to be replaced
by the mass operator $\hat \mu$  in the preconditioned case. Eq.
(\ref{O2}) and  (\ref{ionsdin}) reproduce the adiabatic dynamics of
the ions when the appropriate decoupling condition, Eq.
(\ref{decouple}) discussed in Sec. II, is satisfied \cite{Car,lh}. We
considered the vibrational motion of a Si$_3$ molecule during a time
span of about 0.3 ps. In the unpreconditioned  case we used a time
step $\Delta$=7 a.u. to integrate the equations of motion. This is
close to the maximum allowed time step for a fictitious electronic
mass $\mu$=300 a.u.. Preconditioning allowed to increase this time
step to  $\Delta$ =15 a.u. for a mass  $\mu_0$=260 a.u. and a
preconditioning cut-off $E_p$=1 Ry. In spite of the significantly
larger time step the preconditioned dynamics proceeded adiabatically
in the same way as the one without precontioning. In particular, any
systematic energy transfer from the ionic system to the electronic one
was absent. We plot in Fig. (3) the temporal evolution of the ionic
kinetic energy and of the longest side of the Si$_3$ molecule as  a
function of time in both the preconditioned and the unpreconditioned
cases. Differences between the two dynamics are not noticeable.

\subsection{Damped dynamics in Insulators}

In order to assess the efficiency of the various minimization
dynamics discussed in this work, we considered a 64 atom amorphous
Si sample generated by ab-initio molecular dynamics\cite{Sticha}.
We notice that this system
has a finite gap, and therefore a non-zero $\om_{min}$.
In all our total energy
minimizations we used the same set of starting trial wavefunctions.
These were obtained by minimizing the total energy with a very small
energy cut-off $E_{cut}$ of 2 Ry. We then minimized the total energy
with a cut-off of 12 Ry using four types of dynamics, namely steepest
descent  and second order damped dynamics both without and with
optimal preconditioning. We report the results in Fig. (4). In
particular, we found that, when using the optimal value $\gamma_{opt}$
of Eq. (\ref{gammopt}), the rate of convergence of second order damped
dynamics is faster than that of steepest descent dynamics by the
amount expected from the theoretical analysis in Sec. III.
Preconditioning accelerated further the rate of convergence, so that
finally the rate of convergence of preconditioned second order damped
dynamics was 14 times faster than the one of unpreconditioned steepest
descent.

We determined the value $\gamma_{opt}$  by a rough estimate of $\bar
K_{min}$ based on steepest descent dynamics. In particular,  a 3 point
fit of the exponential decay of the total energy gives:
\be \gamma_{opt}\Delta\sim\sqrt{\frac{1}{2}log
\left(\frac{E_1-E_2}{E_2-E_3} \right)}\label{optgamma}\ee
where $E_1$ $E_2$ $E_3$ are the energies at three successive steps of
steepest descent. We waited until only the slowest exponential was
left in the decay. If faster exponential are still present, Eq.
(\ref{optgamma}) overestimates $\gamma_{opt}\Delta$. In a practical
calculation, we therefore suggest to start the minimization with a few
steps of steepest descent and to use Eq. (\ref{optgamma}) to obtain an
upper bound for the optimal $\gamma\Delta$. Then, we suggest to
proceed with the damped second order minimization, readjusting
$\gamma\Delta$ in order to achieve the optimal limit of critical
damping. As we can see from Fig. (4), it is indeed convenient to use
steepest descent in the first steps of minimization when the highest
frequency components dominate the deviation of the energy from the
minimum. Subsequently, when only the slowest frequencies are left,
damped dynamics becomes much more convenient, especially in those
cases of utterly slow convergence rate.

\subsection{Damped dynamics in Metallic Systems}

In order to test the efficiency of our damped dynamical scheme for minimization
in the case of metallic systems
we applied it to liquid silicon which is a metal. We use a 64 atom sample
generated by ab-initio molecular dynamics\cite{Sugino}.
As explained in Sec. IV, the damping constant $\gamma$ can be fixed
on the basis of the required energy resolution $E_{err}$, for which
we chose here a value of about $20meV$.
We minimized the total energy
with a cutoff of 12 Ry using four types of dynamics, similarly to what we
did in the insulating case. Again the starting trial wavefunctions
were obtained by a minimization using a small cutoff of 2 Ry.
We report the results in Fig. (5). Notice that in the present
metallic case,  steepest descent dynamics is particularly
inefficient, while damped dynamics is very effective, since it
improves by many orders of magnitude the convergence rate of steepest descent.
A further gain results from preconditioning.

\subsection{Comparison with Conjugate Gradient Minimization}

In this subsection we compare our damped dynamical method with a
conjugate gradient minimization scheme.
The standard conjugate gradient procedure, described e.g.
in Ref.~(\onlinecite{numerical}),
cannot be directly applied to a constrained functional, unless
some additional simplifying assumptions are invoked which can reduce
the minimization efficiency \cite{Stich,Teter}. To fully exploit the power of
the conjugate gradient procedure the authors of Ref.~(\onlinecite{Arias})
proposed to use
an unconstrained energy functional. We adopt the same procedure of
Ref.~(\onlinecite{Arias})
but we use a different form for the unconstrained energy functional.
We use
the form suggested in Refs. (\onlinecite{Mauri,Mauri2}) in the context
of electronic structure calculations
with linear size scaling without imposing any localization constraints on
the electronic orbitals\cite{nota linear}.
For reasons of numerical simplicity we adopt here a total energy functional
based on a non self-consistent tight-binding Hamiltonian.
This choice simplifies considerably the line minimization in the
conjugate gradient scheme, which can be performed exactly \cite{Mauri2}.

We used a tight-binding Hamiltonian for carbon \cite{Wang}, and we considered
an ionic liquid configuration
of 64 atoms at a temperature of 5000~K.
For this configuration the system is metallic.
Our results are reported in Fig. (6) where we plot
the logarithmic error in the total energy per atom
versus the number of iterations for various minimization schemes, namely
damped dynamics, conjugate gradient and steepest descent
minimization. In the case of conjugate gradient minimization the number of
iterations was multiplied by a factor of 2 in order
to take into account the increase in computational cost
arising from line minimization.
{}From Fig. (6) it is evident that the numerical efficiency of
both conjugate gradient and of damped molecular
dynamics is considerably superior to that of steepest descent minimization.
In the present example the numerical efficiency of conjugate gradient and
that of damped molecular dynamics minimization are practically the same.

We expect that the results that we have found here should
remain valid also of in the case of a self-consistent LDA Hamiltonian.

\subsection{Charge sloshing on very long cells}

In order to study charge sloshing effects we used a tetragonal supercell
having a long side. In particular we considered crystalline silicon in the
diamond structure and we constructed two supercells by repeating four or
eight elementary cubic cells along the crystallographic (100) direction.
The resulting supercells contain 32 and 64 atoms respectively.
Then we broke the translational invariance of the diamond lattice by giving
the silicon atoms a random displacement of about 5 percent of the bond length.
This does not modify the insulating character of the system.

In the present example we have considered only preconditioned steepest descent
minimization.
As in the previous subsections we prepared the initial trial state by
minimizing the total energy with a small cutoff of 2 Ry starting from a set
of random wavefunctions. Severe charge sloshing instabilities immediately
showed up during this initial minimization in which the starting random
wavefunctions were very far from the converged insulating ground-state.
In particular, already for the 32 atom cell the time step for numerical
integration had to be reduced by an order of magnitude compared to the time
step that we could use in an equivalent situation with a smaller cell.
Such instability was completely eliminated by replacing the Coulomb by a
Yukawa potential as described in Sec. IV. We adopted here a parameter
$2\pi/\alpha=20.5$ a.u. for the Yukawa potential.
Once obtained the initial trial state, we performed a total energy minimization
on the 32 and on the 64 atom cell with a cutoff of 12 Ry.
The results are shown in Fig. (7) which reports the deviation from the
converged ground-state energy as a function of the number of numerical time
steps. During the initial 30 steps we used the Yukawa potential. This allowed
us to use for both 32 and 64 atom cells
an integration time step equal to the one usually adopted for the same system
when using cells sufficiently small that no charge sloshing effects are
present. Then we switched from Yukawa to Coulomb potential.
After 30 minimization steps with the Yukawa potential the system was insulating
and already very close to its exact ground-state.
In this case, as shown analytically in Sec. IV charge sloshing
instabilities are not expected to occur in a periodic system. Indeed
not even in our disordered sample they did occur.
Therefore we could use
in the final 30 minimization steps the same time step used with the Yukawa
potential. The overall convergence rate, as it can be clearly seen
in the figure, is independent from the cell size.

\section{Conclusions}

We have presented a detailed analysis of the stability and of the convergence
rate of fictitious dynamical methods for electrons. We have
succeeded in improving considerably the efficiency of currently used
algorithms for total energy minimization and for ab-initio molecular dynamics.

In the case of ab-initio molecular dynamics simulations
of the ionic motion we have introduced a novel preconditioning scheme which
gives rise to an overall saving of CPU time of the order of 2-3 in typical
applications. In the case of total energy minimization we have introduced
an optimal damped preconditioned dynamics which has a convergence rate
substantially faster than steepest descent algorithms and comparable to
that of the best conjugate gradient schemes for electronic structure
calculations. This is especially important in metallic situations.

Although in this paper we confine our analysis to electronic minimization,
we stress that
the damped dynamics algorithm can also be applied to
ionic minimization. In this case the optimal ionic damping parameter is related
to the phonon frequencies of the system under study.

In addition, we have presented a detailed analysis of the charge sloshing
instability and we have indicated a practical way to control it.
We have shown with a numerical example that, in the case of insulators,
our approach
allows us to converge to the ground-state with a number of iterations that is
independent of the system size.

\acknowledgements{We gratefully acknowledge many useful discussions with
Giulia Galli and Osamu Sugino. We acknowledge support
by the Swiss National Science Foundation
under Grant No. 21-31144.91.}

\section{Appendix}

In this appendix we show that for a periodic insulator the function S(p)
given in Eq. (\ref{S(p)}) goes to zero like $p^2$ for $\bf p$ going to zero.
\be
\begin{array}{rcl}
S({\bf p})&=&{1\over \Omega}\sum_{ik}< \chi^0_i |e^{-i{\bf p r}}| \chi^0_k >
< \chi^0_k |e^{+i{\bf p r}}| \chi^0_i>\nonumber\\
&=&
{1\over \Omega}\sum_{i}(1-\sum_j< \chi^0_i |e^{-i{\bf p r}}|\chi^0_j >
< \chi^0_j |e^{+i{\bf p r}}| \chi^0_i>),\label{a1}
\end{array}
\ee
where the indices $i$ and $j$ refer to occupied states and the index $k$
refers to empty states.
In Eq. (\ref{a1}) we used the completeness relation:
$\sum_k| \chi^0_k >
< \chi^0_k | = {\bf 1} -\sum_j|\chi^0_j >
< \chi^0_j |$.
Let us suppose for simplicity that we have a single occupied band.
Since the expression in Eq. (\ref{a1})
is invariant under unitary transformations on the
occupied subspace, we can write it in terms of Wannier functions, i.e.:
\be
S({\bf p})={1\over \Omega_{min}}
( 1-\sum_{\bf R}|<W_{\bf 0}|e^{-i{\bf p r}}|W_{\bf R}>|^2),\label{a2}
\ee
where $W_{\bf R}$ is the Wannier function centered on site $\bf R$,
and $\Omega_{min}$ is the
volume of the elementary cell. The Wannier functions
are exponentially localized in the case of an insulator: this allows us to
expand in a Taylor series for $\bf p$ going to zero the exponentials
in Eq. (\ref{a2}).
In particular  if we consider the term with ${\bf R}={\bf 0}$ in Eq.
(\ref{a2}),
and  expand the  exponentials in $\bf p r $
around $ {\bf p} <{\bf r}> = {\bf p} <W_{\bf 0}|{\bf r}|W_{\bf 0}>$, we get:
\be
\begin{array}{l}
 1-|<W_{\bf 0}|e^{-i{\bf p r}}|W_{\bf 0}>|^2 =\nonumber\\
\phantom{xxxxx}= 1 -   |<W_{\bf 0}|  {\bf 1} -  i{\bf p}({\bf r}-<{\bf r}>)
- [ {\bf p}({\bf r}-<{\bf r}>)]^2/2|W_{\bf 0}>|^2+o(p^2)\\
\phantom{xxxxx}=
+<W_{\bf 0}|[ {\bf p}({\bf r}-<{\bf r}>)]^2|W_{\bf 0}>+o(p^2)\nonumber
\end{array}
\ee
This term tends to zero as $p^2$.
In a similar way one can show that the terms
with $\bf R$ different from zero in Eq. (\ref{a2}) also go to zero as $p^2$.

\begin{figure}
$\bar \om_{max}/\bar \om_{min}$ as a function of
the preconditioning cut-off $E_p$ for the Si$_3$ molecule.
A periodically repeated cubic cell of 20 a.u. is used in all the calculations
for the Si$_3$ molecule.
\label{fig1}
\end{figure}

\begin{figure}
Spectra of the electronic frequencies for the Si$_3$ molecule.
The solid line refers to 2nd order dynamics without preconditioning.
The dashed line refers to 2nd order
dynamics with preconditioning ($E_p$=1 Ry).
\label{fig2}
\end{figure}

\begin{figure}
Ionic dynamics of the Si$_3$ molecule without
(solid line) and with (dots) preconditioning. In (a) we report the
oscillations of the long side of the Si$_3$ molecule, and in (b) the
oscillations of the ionic kinetic energy as  a function of
time. \label{fig3}
\end{figure}

\begin{figure}
Total energy minimization for a 64 atom amorphous silicon sample,
using non-preconditioned steepest descent
(SD NP), preconditioned steepest descent (SD P), non-preconditioned
damped dynamics (D NP) and preconditioned damped dynamics (D P). We
plot the logarithm of the difference between the energy per atom ($E$)
and the ground state energy per atom ($E_0$) in Hartree units
vs. the number of integration steps.\label{fig4}
\end{figure}

\begin{figure}
Total energy minimization for a 64 atom liquid silicon sample,
using non-preconditioned steepest descent
(SD NP), preconditioned steepest descent (SD P), non-preconditioned
damped dynamics (D NP) and preconditioned damped dynamics (D P). We
plot the logarithm of the difference between the energy per atom ($E$)
and the ground state energy per atom ($E_0$) in Hartree units
vs. the number of integration steps.\label{fig5}
\end{figure}

\begin{figure}
Total energy minimization for a 64 atom liquid carbon sample,
using steepest descent
(SD), conjugate gradients (CG) and
damped dynamics (D). The total energy corresponds to a parameterized
tight-binding Hamiltonian (see text). We
plot the logarithm of the difference between the energy per atom ($E$)
and the ground state energy per atom ($E_0$) in Hartree units
vs. the number of integration steps.
The number of integration steps of the
conjugate gradient calculation has been multiplied by two to take into account
the increase in computational cost compared to the other methods.
\label{fig6}
\end{figure}

\begin{figure}
Total energy minimization for two randomized crystalline
silicon samples using preconditioned steepest descent.
The continuous line refers to a 32 atom cell with a long side of 41 a.u.
The dashed line refers to a 64 atom cell with a long side of 82 a.u. We
plot the logarithm of the difference between the energy per atom ($E$)
and the ground state energy per atom ($E_0$) in Hartree units
vs. the number of integration steps. In panel (a) we use a Yukawa potential
to compute $E$ and $E_0$, while in panel (b) the Yukawa potential is replaced
by a Coulomb potential (see text).
\label{fig7}
\end{figure}


\begin{references}
\bibitem{nota linear} Since we do not impose localization constraints
on the electronic
orbitals the cost of the calculation still grows as the cube of the
system size,
as in standard electronic structure calculations. As shown in
Ref.~(\onlinecite{Mauri}) the miminum
of the unconstrained energy functional coincide with that of the constrained
functional. Moreover, since we
adopt a truncation parameter ${\cal N}=1$ (see Ref.~(\onlinecite{Mauri})) for
the
inverse of the overlap matrix, the
minimization rate is the same for the unconstrained
and the constrained functional\cite{Mauri2}.
\end{references}
\end{document}